\documentclass[fleqn,usenatbib]{mnras}

\usepackage[T1]{fontenc}
\usepackage{ae,aecompl}

\usepackage{graphicx}	
\usepackage{amsmath}	
\usepackage{amssymb}	

\usepackage{newtxtext,newtxmath}

\title[GK Persei in 2018]{Swift monitoring of GK Persei during the 2018 dwarf nova outburst}

\author[Pei et al.]{
Songpeng Pei,$^{1}$\thanks{E-mail: songpengpei@outlook.com}
Marina Orio,$^{2,3}$
and Xiaowan Zhang$^{1}$
\\
$^{1}$School of Physics and Electrical Engineering, Liupanshui Normal University, Liupanshui, Guizhou, 553004, China\\
$^{2}$INAF-Osservatorio di Padova, vicolo Osservatorio, 5, 35122 Padova, Italy\\
$^{3}$Department of Astronomy, University of Wisconsin, 475 N. Charter Str., Madison, WI, USA\\
}

\date{Accepted XXX. Received YYY; in original form ZZZ}

\pubyear{2015}

\begin{document}
\label{firstpage}
\pagerange{\pageref{firstpage}--\pageref{lastpage}}
\maketitle

\begin{abstract}
The old nova and intermediate polar (IP) GK Persei underwent one of its recurrent dwarf nova (DN) outbursts in 2018. We proposed monitoring it in UV and X-rays with the Neil Gehrels Swift Observatory, starting less than six days after the eruption, until 16 days after the eruption ended. For the first time we could follow the decay to minimum light UV and X-rays.
We present the timing and spectral analysis, comparing the results with the previous outbursts and with the quiescent status. We confirm the spin modulation in X-rays with a period $P_{\rm WD} = 351.325(9)$ s, only in the 2 – 10 keV range. The period was not detected in the 0.3 – 2 keV range and in the UV band, suggesting that the soft portion of the X-ray spectrum in GK Per does not originate
 near the poles, but in a wind or circumstellar material. The amplitude of the modulation was less prominent than in 2015, a fact that seems correlated with a lower average mass accretion rate. The spectral fits are consistent with a mass accretion rate increasing by a factor of 2 from rise to maximum and decreasing during the return to minimum, following the trend of the modulation amplitude. The maximum plasma temperature is higher than the {\sl Swift} XRT energy range of 0.3 – 10 keV, thus it is not well constrained, but our spectral fits indicate that it may have varied irregularly during the outburst.

\end{abstract}




\section{Introduction} \label{sec:intro}

GK Persei (A 0327+43, Nova Persei 1901) is a cataclysmic variable (CV) with unique characteristics. 
At a distance of 441 $_{-6}^{+13}$ pc \citep{2018MNRAS.481.3033S}, it is the second closest nova yet detected.
It is one of the CVs with the longest orbital periods, 1.9968 d \citep{1986ApJ...300..788C,2002MNRAS.329..597M, 2021MNRAS.507.5805A}.
 Its secondary is in fact an evolved star, a carbon-rich K2-type subgiant with mass of 0.25 -- 0.48 M$\odot$ \citep{1985MNRAS.212..917W,2002MNRAS.329..597M, 2021MNRAS.507.5805A}. 
 
In 1901, GK Per underwent one of the brightest nova outbursts ever recorded 
 \citep{1901ApJ....13..170P, 1901ApJ....13..173H, 1901MNRAS..61..337W}. It immediately attracted much attention, not only because of its luminosity, but also because of an intriguing and famous light echo \citep[][]{Ritchey1901, Ritchey1902, Perrine1901}, which was interpreted and discussed over the years in several papers \citep{Kapteyn1901, Couderc1939, Tweedy1995}. A light echo was observed again only twice in stellar explosions (in Nova Sgr 1936 and SN 1987 A). More than a century after the nova eruption,
 GK Per was again "in the news", as an object of press-releases, because it hosts the first classical nova remnant discovered in X-rays \citep{2005ApJ...627..933B}. 
While it was often monitored as a post-nova 
because of its interesting
 peculiarities, it was first observed to undergo a dwarf-nova-like (DN-like) outburst in 1948, with amplitude of 1 -- 3 mag. Similar 
 outbursts, lasting for up to 2 -- 3 months, were then observed to recur steadily at intervals a little longer than 2 years, roughly expressed as n $\times$ $(400\pm40)$ days \citep{1983A&AS...54..393S, 2002A&A...382..910S}.
 
 Furthermore, GK Per has also been studied because it contains a magnetized WD \citep{1983A&A...125..112B} and is classified as an IP \citep[see][]{1985MNRAS.212..917W}. IPs are CVs in which an accretion disk is formed, but it is disrupted by the magnetic field and accretion is funneled to the poles. The long DN outbursts in IPs are attributed to the thermal-viscous instability of the accretion disc \citep{Hameury2017}. We know that the magnetic field of the WD in GK Per has been estimated to be 0.5 Megagauss \citep{2018MNRAS.474.1564W}, and that it seems fairly clear that the DN outburst occurs in the inner part of the accretion disk \citep{1986A&A...160..367B,1992ApJ...384..269K, 1996PASP..108...39O, 2002PASJ...54..987N}. \citet{2009MNRAS.399.1167E} estimated that the accretion disk of GK Persei has a radius of $\sim$ 2 $\times$ 10$^{11}$ cm, while only a small fraction of disk is involved in the instability \citep[see also][]{2018MNRAS.474.1564W}.

Several parameters have been estimated thanks to observations in the X-ray range, which is an important window to understanding the physics of GK Per. The plasma temperature inferred from the X-ray spectrum increases with WD mass, because of a stronger shock where the gravitational potential is higher. \citet{2016A&A...591A..35S} estimated the mass of the WD as $M_{\rm WD} = 0.86~\pm~0.02~M_{\rm \odot}$, and \citet{2018MNRAS.474.1564W} constrained it as 0.87 $\pm$ 0.08 M$\odot$ , by using the temperature and flux measured in hard X-rays with {\sl NuSTAR} and the standard WD mass–radius relation. 

Another essential parameter exactly measured in X-rays in IPs is the spin period of the WD. For GK Per
 X-ray observations also revealed a period of about 351 s \citep{1985MNRAS.212..917W}, and spin-up rate of 0.0003 s yr$^{-1}$ \citep{2017MNRAS.469..476Z}. Rotation periods of the WD in IPs are often of several minutes, and they are a signature of these systems, in which the rotation and orbital periods are not synchronized and the X-ray flux originates close to the poles. Presumably, the X-rays emitted by the shocked plasma impinging the surface of the WD become absorbed by the accretion stream or curtain during the rotation period \citep[see][and references therein]{2017PASP..129f2001M}.
The broadband X-ray spectrum in IPs is usually due to the superposition of a hard and a soft component. The soft X-ray emission is thought to be due to the reprocessing of the hard X-rays on the surface of the accreting poles in the WD atmosphere, and it can be approximated as a blackbody 
with temperatures ranging from 30 eV to 120 eV \citep{1996A&A...310L..25B, 2004NuPhS.132..693D, 2004A&A...415.1009D, 2006A&A...449.1151D, 2006A&A...454..287D, 2009MNRAS.392..630L, 2017PASP..129f2001M}.
With an orbital inclination angle 
$i$ of the binary system in the range of $63 - 73^{\circ}$ \citep{2018MNRAS.474.1564W, 2021MNRAS.507.5805A}, the spin
 period of GK Per is clearly measured. However, a peculiarity of GK Per is that the period is detected only above 2 keV \citep{2017MNRAS.469..476Z}, while in most other IPs it is either only, or mostly, evident in the soft, black-body-like portion of the spectrum, consistently with absorption as root cause of the modulation. This atypical behaviour is found both in outburst and in quiescence \citep{2017MNRAS.469..476Z}; the only difference between the two states seem to be that
 the spin modulation is double peaked in quiescence
\citep{1991PASP..103.1149P, 1992MNRAS.254..647I}, while \citet{2004MNRAS.349..710H} found it instead nearly sinusoidal during one of the DN-like outbursts. 

As the accretion rate increases during the DN outbursts of GK Persei, the magnetosphere is pushed closer to the WD surface. Thus, during outburst, the accretion disk inner radius decreases together with the magnetospheric radius $R_{\rm m}$ \citep{2004MNRAS.349..710H, 2005A&A...439..287V, 2016A&A...591A..35S}. Consistently with this model, the maximum plasma temperature (indicative of the strength of the shock, which
 decreases if the accretion stream to the poles begins from a point closer to the WD) during the DN outbursts has been estimated, by fitting X-ray spectra, to be quite lower than in quiescence \citep{2009A&A...496..121B, 2017MNRAS.469..476Z, 2018MNRAS.474.1564W}. 

In addition to the spin period, in GK Per both aperiodic variability
 and quasi-periodic oscillations (QPOs) with 
characteristic time-scale of several kiloseconds have been discovered in X-rays and optical during the
outburst \citep{1985MNRAS.212..917W, 1994ApJ...424L..57H, 1996MNRAS.283L..58M, 
1999MNRAS.306..753M, 2002PASJ...54..987N, 2004MNRAS.349..710H, 2005A&A...439..287V}.
 QPOs with time-scales of 380 s and 400 s were also discovered, but 
only in the optical \citep{1981ApJS...45..517P, 1985A&A...149..470M}.
 \citet{1994ApJ...424L..57H, 2004MNRAS.349..710H} suggested that an 
often observed 5000 s QPO is caused by bulges in
 the disk, due to overflows as matter is accreted, moving
with the local Keplerian time-scale of about 5000 s at the inner disc radius.

The outburst in 2018 had a large amplitude;
the AAVSO light curve shows that a few outbursts in past years had smaller amplitude and different duration with
 multiple peaks, like, for instance, the one that occurred in 2006. In previous years the nova was too close to the Sun to be observed immediately after optical, UV and X-ray maximum. For the first time in 2018, we could obtain the X-ray and UV light curves in the post-maximum phase, during the decay to quiescence. Because of the many "surprises" of this interesting system, we proposed to monitor the complete course of the 2018 DN-like outburst in X-rays and UV, obtaining UV and X-ray light curves over the whole duration of the outburst.

\section{Observation and data reduction} \label{sec:observation}
The Neil Gehrels Swift Observatory \citep[hereafter, {\em Swift};][]{2004ApJ...611.1005G} observations we proposed for the 2018 outburst of GK Persei began on 2018-08-29, that is 5.55 days after the eruption time of 2018 August 23.5 (AAVSO Alert Notice 649), and ended on 2018-11-12.
During 16 observations the Swift UV/Optical Telescope \citep[UVOT;][]{2005SSRv..120...95R} was in the event mode, providing fast
 photometry in the UVW1 filter, that was
 chosen for comparison with \citet{2009MNRAS.399.1167E} and \citet{2005A&A...439..287V}.
The event mode was used to extract the light curves for timing analysis in comparison with \citet{2009MNRAS.399.1167E}.
Swift UVOT was in the image mode in the other observations, providing the mean magnitude of each observation in one of the three UVOT filters (U, UVW1 and UVW2) for comparison with the previously measured magnitudes in the same filters in 2015. All the Swift UVOT data were processed by using the FTOOLS package.

The Swift X-ray Telescope \citep[XRT;][]{2005SSRv..120..165B} monitoring was performed only in the photon counting (PC) mode. 
 The XRT light curves and spectra were obtained with the XSELECT tool of FTOOLS package in HEASOFT v. 6.30.1\footnote{https://heasarc.gsfc.nasa.gov/docs/software/lheasoft/download.html}. The central region was excluded, and the XRTLCCORR command was used to correct the effect of pile-up. The XRT light curves were barycentric corrected by using the BARYCORR tool. The processed Swift Burst Alert Telescope \citep[BAT;][]{2005SSRv..120..143B} data obtained from the Swift BAT transient monitor page was also used \citep{2013ApJS..209...14K}. XSPEC v. 12.12.1 \citep{1996ASPC..101...17A, 2003HEAD....7.2210D} was used to analyze and fit the X-ray spectra.

The list of all the observations with the date, exposure time and mean count rate is presented in Table~\ref{tab:obsgkper}.

\begin{table}
\begin{minipage}{80mm}
\caption[Log of all observations of Swift XRT (PC mode)]{Log of all observations of Swift XRT (PC mode) of GK Persei during the 2018 outburst.}
\label{tab:obsgkper}
\begin{tabular}{lccc}
\hline
\hline

      ObsID& Date$^{a}$ & Exp. (s)& Count rate  \\
\hline
 00010873001&58359.05  &163.0  & 1.25 $\pm$ 0.11  \\
00010873002 &58359.06  &1454.2  & 1.23 $\pm$ 0.04  \\
00010873004 &58359.19  &1471.8  &0.57  $\pm$ 0.02 \\
00030842076 &58363.17  & 1750.1 &0.74  $\pm$ 0.04 \\
00030842077 &58367.71  & 1153.4 &1.92  $\pm$ 0.07 \\
 00030842078& 58370.09 &1479.3  &1.05  $\pm$ 0.05  \\
00030842079 &58375.53  & 1830.3 & 0.86 $\pm$ 0.04 \\
00030842080 & 58378.78 & 997.9 & 1.50 $\pm$ 0.05 \\
 00030842081&58381.19  &1063.1  & 0.70 $\pm$ 0.04  \\
00030842082 &58384.75  &887.6  &1.36  $\pm$ 0.05  \\
00030842083 &58387.50  & 1018.0 &1.19  $\pm$ 0.04 \\
00010907001 & 58396.38 & 122.9 &0.68  $\pm$ 0.08  \\
00010907002 & 58396.39 & 574.2 &0.87  $\pm$ 0.05  \\
00010907003 & 58396.59 &168.0  &0.51  $\pm$ 0.07  \\
00010907004 &58396.59  &576.7  &0.96  $\pm$ 0.06  \\
00030842084 &58399.38  &210.6  & 0.85 $\pm$ 0.07  \\
00030842085 &58402.55  &1973.2  & 1.75 $\pm$ 0.04  \\
00030842086 & 58405.49 &423.7  &1.41  $\pm$ 0.07 \\
00030842087 &58411.19  &1697.4  &1.34  $\pm$ 0.04 \\
00010944001 & 58415.91 &115.3  &0.90  $\pm$ 0.09 \\
00010944002 &58415.91  &1113.2  & 0.65 $\pm$ 0.04 \\
00010944003 & 58415.96 &110.3  & 0.47 $\pm$ 0.07 \\
00010944004 &58415.97  & 1175.9 &0.85  $\pm$ 0.04 \\
00030842088 &58422.34  &895.1  & 0.23 $\pm$ 0.02 \\
00030842089 &58426.74  & 945.3 & 0.33 $\pm$ 0.02  \\
 00030842090&58430.85  &852.5  & 0.37 $\pm$ 0.03 \\
00030842091 & 58434.90 & 922.7 & 0.51 $\pm$ 0.02 \\
\hline

\multicolumn{4}{p{.9\textwidth}}{{\bf Notes: }$^a$ Modified Julian Date.
The count rates were measured in the 0.3 -- 10.0 keV.}\\

\end{tabular}
\end{minipage} 
\end{table}

\section{Timing analysis}

\begin{figure*}\begin{center}
\includegraphics[height=15.2cm]{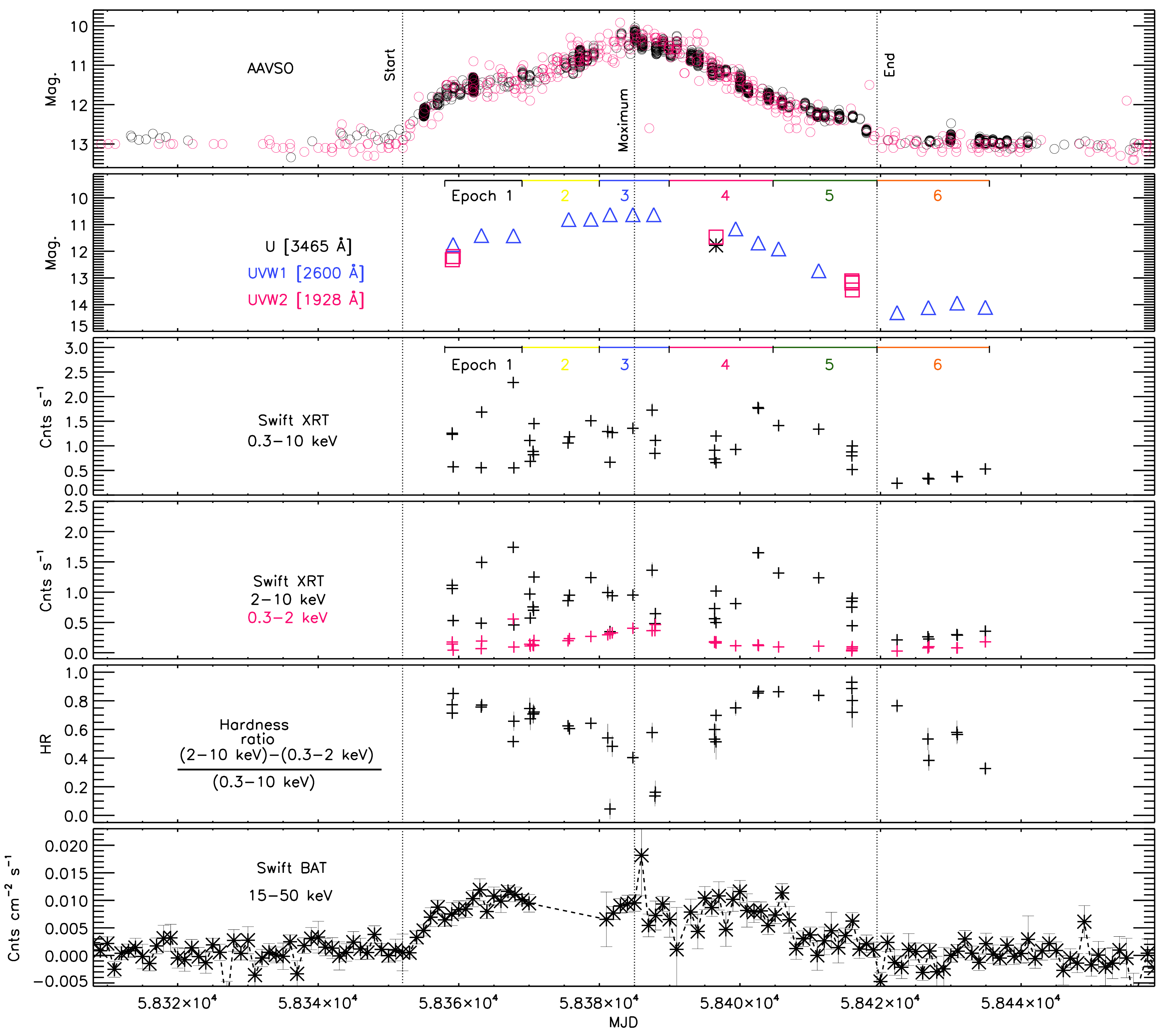}
\end{center}
\caption[Light curves in different filters]{From top to bottom: AAVSO light curve of GK Persei in the V band (black) and without a filter (red). The three vertical dotted lines in all the panels mark the beginning (MJD 58353.5), the maximum (MJD 58384.9) and the end (MJD 58419.5) of the 2018 outburst in the optical band. The next panels show the Swift UVOT light curves in different filters, the Swift XRT light curve in 0.3 -- 10 keV in the PC mode, the Swift XRT light curves in 2 -- 10 keV (black) and 0.3 -- 2 keV (red), the X-ray hardness ratio: ((2 -- 10 keV) -- (0.3 -- 2 keV))/(0.3 -- 10 keV), and the Swift BAT light curve. The horizontal lines with number labels in the second and third panels represent six epochs defined for the analysis.}
\label{fig:new hr}
\end{figure*}

The comparison between the development of the 2018 dwarf nova outburst in the optical, UV and X-rays in different energy regions is shown in Fig.~\ref{fig:new hr}. In order to study the
temporal evolution of the spin profiles and spectral shapes, the data were divided into six time intervals, labelled in Fig.~\ref{fig:new hr} as Epoch 1 to Epoch 6. Epoch 3 was chosen at the time around the maximum of the outburst in the optical, which is also to the X-ray maximum. The period after the end of the outburst in the optical was defined as Epoch 6.

We used the Lomb-Scargle periodograms (LSPs) method \citep{1982ApJ...263..835S} to search the periods in the soft and hard energy bands light curves of GK Persei. Specifically, we used the Package ‘lomb’, written by Thomas Ruf\footnote{https://cran.r-project.org/web/packages/lomb/index.html}. This package computes the Lomb-Scargle periodogram (LSP) for unevenly sampled times series (e.g., series with missing data). The p-value for the significance of the largest peak in the periodogram is computed from the exponential distribution. We set the false alarm probability level at 0.3\%. When it was necessary to additionally check the significance of the periodicity we also repeatedly scrambled the data 100000 times and calculated the probability that random peaks in LSP exceeded the height of the main peak in the original LSP. The 1 sigma error for the searched period was estimated by using the Gaussian model to fit the corresponding peak in the LSP.

\begin{figure*}
\begin{center}
\includegraphics[width=7.1cm]{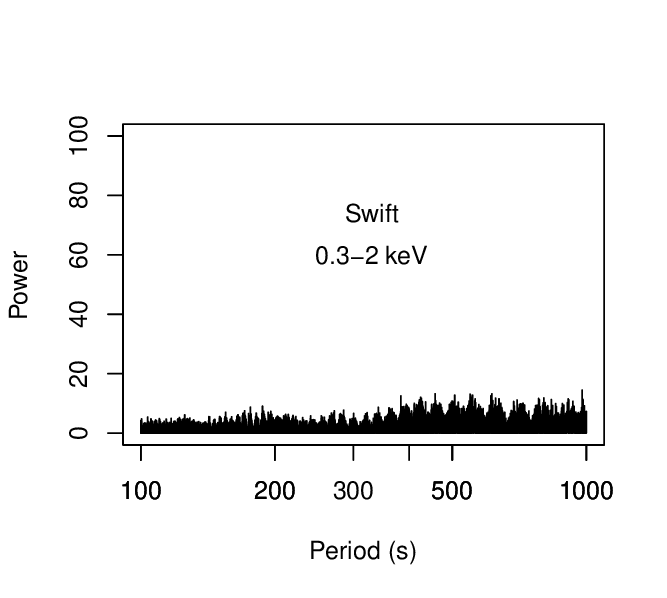}
\hspace{2.62em}
\includegraphics[width=7.1cm]{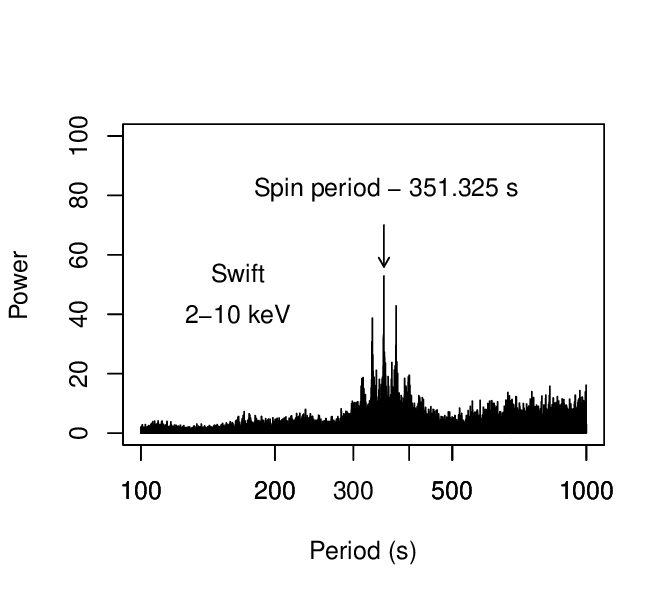}
\caption[The LSP plots]{Left: the LSP obtained from the 20 seconds per bin light curve of the Swift XRT data in the 0.3 -- 2 keV energy range. Right: the LSP obtained from the 20 seconds per bin light curve of the Swift XRT data in the 2 -- 10 keV energy range.}
\label{fig:lsp}
\end{center}
\end{figure*}

\begin{figure*} 
\includegraphics[width=15cm]{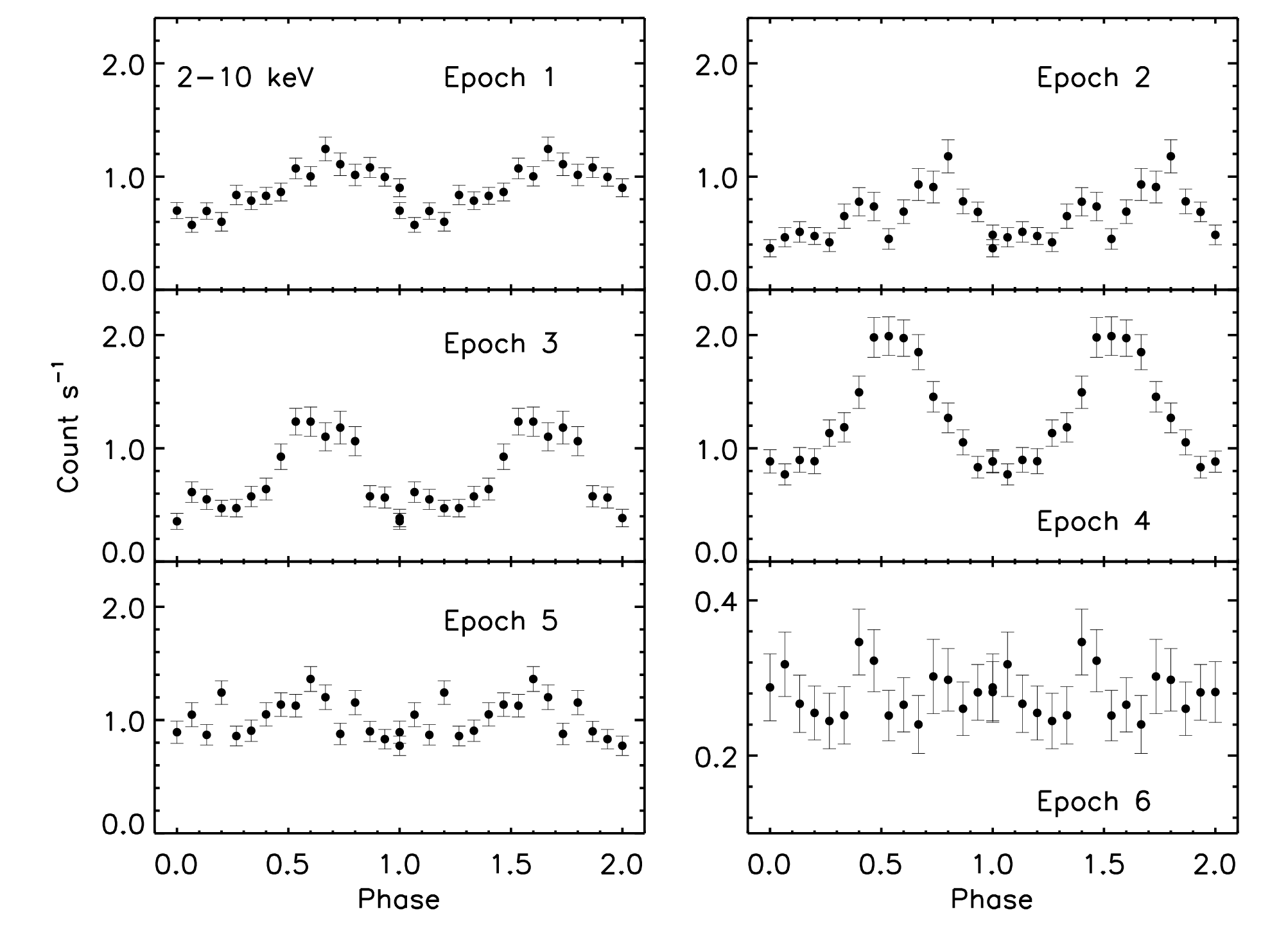}
\caption[Time evolution of spin-folded light curves]{Time evolution of spin-folded light curves of the 2018 data of GK Persei in 2 -- 10 keV. From top to bottom, the six panels represent light curves for Epoch 1 -- 6. In each panel, the abscissa covers two spin phases for better visibility.}
\label{fig:pfold}
\end{figure*}

We analysed the light curve in the 0.3-2 keV range, and in the 2-10 keV range separately, to compare the results with \citet{2017MNRAS.469..476Z}. In the 2015 outburst, in fact, the spin period could be detected only at energy above 2 keV, a fact that is rather exceptional for IPs. We also experimented by analysing lightcurves binned with different time bins, specifically 10 s and 20 s. In the periodogram obtained with a 20 s bin there were some peaks that differed from the periodogram obtained with the 10 s bin lightcurve, at periods above 1000 s. We found that only the peak at $ 351.325 \pm 0.009$ s in the 2-10 keV lightcurve is significant, stable and
 not varying with the bin size used in the analysis.
 It is the well known period attributed to the spin of the WD, and it is not 
 measured in the 0.3 $-$ 2 keV light curve. The LSPs for the two energy bands are shown in Fig.~\ref{fig:lsp}, while Fig.~\ref{fig:pfold} shows the spin-folded light curves of each of the 6 epochs defined above, in the 2 $-$ 10 keV energy band. In the folding process, a spin period of 351.325 seconds was used, and the phase 0 epoch was set at 2018-08-27 00:00:00.000 UTC (MJD 58357.0). We see in Fig.~\ref{fig:pfold} 
 that 
(with exposure times lasting between 2968.7 s and 5992.5 s), separately, the period due to the spin is retrieved also in the single epochs. The period modulation amplitude , quantified as in \citet{2009MNRAS.399.1167E}, $(max-min)/(max+min)$, from epoch 1 to 6 is 56\%, 70\%, 74\%, 67\%, 27\% and 21\%, respectively. We note that the changes in the period are not detectable, since the results are within each other's errors.

For the soft X-ray light curve in the 0.2 -- 3 keV range, we are only able to obtain upper limits on the amplitude of the modulation. \citet{Pavlov2009} derived a formula for the minimum detectable amplitude of a periodic modulation in a frequency range $\leq$ {\sl f}, with a given 3 $\sigma$ false detection probability P,
if $N_S$ and $N_B$ are the number of counts from the source and background, respectively. 

\begin{equation}
p \approx \frac{[4ln(t_{exp} * f /P) (N_{S}+N_{B}) ]^{1/2}}{N_{S}} ,
\label{time}
\end{equation}

In the whole exposure time of 25146 s including
 all the epochs, the average count rate did not vary much in any single exposure and was on average over the whole outburst of 0.16$\pm$0.02 cts s$^{-1}$. If we assume P = 0.03,
we obtain a minimum measurable amplitude for a frequency 
 of 1/351 Hz of about 9\%, much
 smaller than we measured in the 2 -- 10 keV band, even for the single epochs.

 It is also important to mention that
we also did not retrieve this period in the light curve of the UV band by using the LSPs method, and we also assessed that the spin period was not present in the UV in any single epoch chosen as above, in the single exposures, nor in any consecutive 2 or 3 randomly chosen exposures.

 We did not expect we would be able to measure with the Swift XRT the $\sim$ 5000 s period found in X-rays (and sometimes also in optical), 
during previous outbursts \citep{1996MNRAS.283L..58M, 2002PASJ...54..987N,
1985MNRAS.212..917W, 1994ApJ...424L..57H, 2005A&A...439..287V}, 
because this period is too close to the 5754 s orbital period of Swift.
The other peaks in fact are all around this orbital period 
or half of it (2877.0 s) and are not energy independent.
 These additional peaks also drift with the counts per bin in the light curves. Because of the 
 fluctuation of the periods with the adopted binning factor and the secondary peaks around the highest one, we definitely attributed 
these periodicities to the Swift orbit, and assessed that they
 are not intrinsic to the source.

\section{Spectral analysis}
The X-ray spectrum of GK Persei is complex. The timing analysis 
 already indicates that there are at least two different sources of X-ray emission in GK Persei, emitting in the hard (2.0 -- 10 keV) and in
 the soft X-rays (0.3 -- 2.0 keV). The observed modulation implies that the hard X-rays are emitted in a region that is obscured during the rotation period, presumably close to the WD poles towards which accretion is funneled.

\begin{figure*}
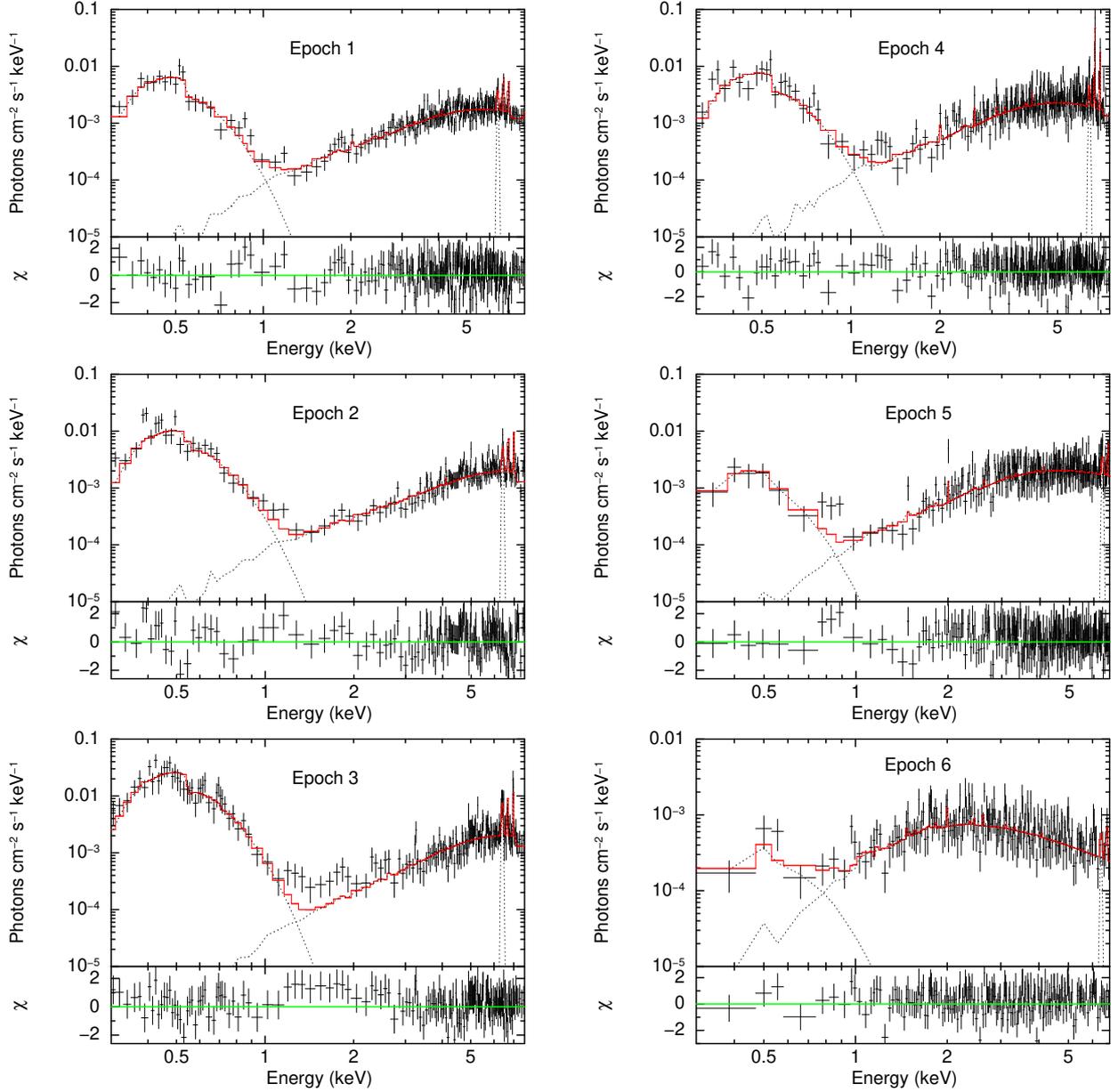

\begin{center}
\includegraphics[width=5.4cm,angle=270,clip]{epoch1.eps}
\hspace{1.62em}
\includegraphics[width=5.4cm,angle=270,clip]{epoch4.eps}
\includegraphics[width=5.4cm,angle=270,clip]{epoch2.eps}
\hspace{1.62em}
\includegraphics[width=5.4cm,angle=270,clip]{epoch5.eps}
\includegraphics[width=5.4cm,angle=270,clip]{epoch3.eps}
\hspace{1.62em}
\includegraphics[width=5.4cm,angle=270,clip]{epoch6.eps}
\caption[Unfolded spectra and the corresponding $\Delta \chi$ of the six epochs]{Unfolded spectra and the corresponding $\Delta \chi$ of the six epochs of GK Persei during the 2018 outburst, using the XSPEC composite model \texttt{tbabs $\times$ (BB + pwab $\times$ (mkcflow + gaussian))}.}
\label{fig: Unfolded spectra}
\end{center}
\end{figure*}

The spectrum above 2 keV can be well fitted with the cooling flow model \citep{1988ASIC..229...53M} (MKCFLOW model in XSPEC), while following the solution of most other authors, including \citet{2017MNRAS.469..476Z}, the portion of the spectrum in the 0.3 -- 1.0 keV range can be fitted with a
blackbody (BB, or BBODY model in XSPEC). Abundances of individual elements are not well constrained, therefore we did not use the version for the cooling flow model that includes varying abundances.
 In the cooling flow model we fixed the lower temperature of the MKCFLOW model at the lowest value of 0.0808 keV in XSPEC, in order to reduce the number of free parameters (the other parameters are not very sensitive to the choice of the minimum temperature).
 In all the six epochs, there is a significant emission feature
 at about 6.4 keV (see Fig.~\ref{fig: Unfolded spectra}), 
with no significant shift in energy; it is the Fe K$\alpha$ fluorescent line at 6.4 keV. A Gaussian component was added to fit this line, with
fixed central energy and line width of the Gaussian at 6.4 keV and 0.04 keV, respectively. 
The Tuebingen-Boulder interstellar medium (ISM) absorption (TBABS) model \citep{2000ApJ...542..914W} (TBABS
in the XSPEC) was used to calculate the cross section for X-ray absorption by the ISM. Following \citet{2017MNRAS.469..476Z} we added to the column density N(H), modeled with power-law distribution of neutral absorbers model (PWAB in the XSPEC), a power-law distribution of covering fraction as a function of column density, built from the wabs code \citep{Done1998}. This addition is necessary for the MKCFLOW+Gaussian portion of the spectrum; we did not obtain a statistically good fit without it. The spectra of Epoch 1 to 6 and the corresponding best fit are shown in Fig.~\ref{fig: Unfolded spectra}, and the best fit parameters are shown in Table~\ref{tab:model parameters}. We also experimented by adding an astrophysical plasma emission code (APEC) \citep{2001ApJ...556L..91S} component like in \citet{2017MNRAS.469..476Z} did to better fit the high resolution spectrum, but this did not improve the fit in our case (\citet{2017MNRAS.469..476Z} also fitted a high resolution Chandra spectrum and measured emission lines, which we did not have). The APEC model (APEC in the XSPEC) represents the
 emission spectrum of an optically thin plasma that is in collisional ionization equilibrium (CIE). We also tried to add another MKCFLOW or BBODY to the \texttt{tbabs $\times$ (BB + pwab $\times$ (mkcflow + gaussian))} model, but this
did not improve the fit. 

An interesting parameter of the MKCFLOW model is the mass accretion rate.
In our fits, this parameter increased by a factor of 2 from rise to maximum, and 
decreased again by a factor of 23 from maximum to minimum. Even if this parameter is of course dependent on other fit parameters, this is a significant result.
%
\section{Discussion}
Fig.~\ref{fig:new hr} shows that the soft X-ray, UV and optical light curves have similar profiles, without any offset in the maximum.
Although there were some irregular variations in the X-ray count rate above 2 keV, the hardness ratio clearly decreased at maximum light, as the flux of the hard component increased. A similar trend was observed in the rising portion of the light curve in the previous outbursts \citep{2015A&A...575A..65S, 2017MNRAS.469..476Z}.

Like in the 2015 outburst \citep{2017MNRAS.469..476Z}, the $\sim$ 351 s period in our data was detected in the 2 -- 10 keV energy range, 
but no modulation with amplitude larger than 9\% was measurable
 in the 0.3 -- 2 keV energy range, indicating a different
origin of the emission in the two energy bands. In 2015, {\sl NuSTAR} observations showed that the modulation was observed even at very high energy, in the 10 -- 40 keV band. \citet{2017MNRAS.469..476Z} suggested that
a greater height and a wider extension of the accretion column in GK Per than in other IPs with shorter orbital separation
 cause only a very moderate energy dependence due to absorption. However, in the 2 -- 10 keV flux we observed an amplitude of the modulation varying between 21\% and 74\%, much larger than the minimum amplitude of 9\% that would have been measurable in the 0.3 -- 2 keV range.
Therefore, even a very moderate inverse energy
 dependence of the modulation is ruled out; we conclude that the softest portion of the spectrum is not modulated with the spin period. Thus, unlike in most known short-orbital-period IPs, here the soft flux most likely does not
 originate in the material that is accreted onto the WD. It is more likely to be due to the circumstellar material of this nova. Although the temperature (T$_{\rm BB}$) of the blackbody component resulting from our fit, in the 65 -- 83 eV range, is consistent with the values derived for the previous outbursts
\citep{2005A&A...439..287V, 2007ApJ...663.1277E, 2009MNRAS.392..630L, 2009MNRAS.399.1167E,2017MNRAS.469..476Z}, the blackbody is only a rough approximation because with the XRT we cannot resolve the emission lines of an ionized plasma. Thus, this component may not be a blackbody at all, but it is rather likely to be ionized material emitting an emission line spectrum.

 \citet{2017MNRAS.469..476Z} discussed the {\sl Chandra} High Energy Transmission Grating (HETG) spectrum of GK Per in the 0.4 -- 10 keV range. At energy 0.7 -- 2.0 keV, the Chandra grating spectrum shows only emission lines. The above authors fitted the soft portion of the spectrum with a blackbody at a temperature of 66 eV, in addition to a thermal plasma component with temperature in the 0.9-5.3 keV range accounting for the emission lines. However, we note that
they could measure the periodic modulation only in the iron lines around 7 keV, and there was no modulation in all the other prominent emission lines. The HETG is not sensitive in the 0.2 -- 0.4 keV range, and there are no archival high resolution X-ray spectra of GK Per at this soft energy, so the nature of the emisison cannot be definitely understood. However, because of this puzzling lack of modulation we suggest that when we extract the X-ray spectrum from the point source observed in the XRT, we are including some flux of a second component of diffuse thermal plasma at low temperature, possibly originating from a disk wind. 
 
 During the whole course of the outburst, we confirmed that the amplitude of the modulation in the 2.0 -- 10 keV light curve was larger than in quiescence, and although our measurements are consistent with those of an outburst in 2006 \citep{2009MNRAS.399.1167E}, the average amplitude of the modulation was definitely smaller
 in 2018 than in the 2015. Our spectral fits, shown in Table~\ref{tab:model parameters}, indicate that the mass accretion rate 
 was in the range 0.7 -- 2.3 $\times$ 10$^{-9}$ M$_\odot$ yr$^{-1}$ during the 2018 outburst, while in 2015, at least during the rise to maximum, the spectral fits yielded much larger values, in the range of 0.6 -- 2.6 $\times$ 10$^{-8}$ M$_\odot$ yr$^{-1}$ (Zemko et al.
 2017). These two facts seem to indicate a positive correlation of the pulsed fraction in outburst with the mass accretion rate. We also note that, in our data, the amplitude of the modulation increased towards maximum, and decreased again during the decay. At minimum, when the modulation amplitude was the smallest, our fits give the lowest estimate of mass accretion rate, only 10$^{-9}$ M$_\odot$ yr$^{-1}$. To summarise, the pulsed fraction appears to be clearly correlated with the mass accretion rate parameter yielded by the spectral fits. 

A characteristic that appears to vary from outburst to outburst is instead the UV periodic variability. Like in 2015 and in 2018, the spin modulation was found to be absent in another outburst in 2002 \citep{2005A&A...439..287V}, but \citet{2009MNRAS.399.1167E} found weak evidence of it in the Swift-UVOT data of 2006. 
We note that the outbursts of 2002, 2015 and 2018 were all ``typical'', namely similar to most other outbursts of GK Per, but the eruption
differed significantly at optical and UV wavelengths 
in 2006, with an optical maximum $\sim$ 1.5 mag fainter than in most other eruptions, and three ``humps'' rather than one. It is tempting to speculate that the detection of the rotation period in the UV in 2006 is connected with different physics at work in the smaller, multi-humped outbursts. However, only new measurements in another such outburst of smaller amplitude, if it happens, may reveal the connection with the physical parameters.

The hydrogen column density of the total absorber is $\sim$ 0.17 $\times$ 10$^{22}$ cm$^{-2}$, consistent with interstellar absorption. The partial absorber, thought to be intrinsic in the binary system, was used in the fit of 9 IPs in outburst, reaching a value of 10$^{23}$ cm$^{-2}$ \citep{2012A&A...542A..22B}. Our fit indicates a similar value for GK Persei in outburst. 

Although we cannot precisely estimate the maximum plasma temperature, because it is above the range of the Swift XRT, from the spectral fits to the Swift spectra during the outburst we obtained values from 20 $_{-4}^{+5}$ to 41 $_{-10}^{+8}$ keV. Thus, the temperature did not appear to evolve in correlation or anti-correlation with other parameters such as the mass accretion rate. We note that in quiescence,
 estimates of the maximum plasma temperature over the years have been: 25 -- 42 keV \citep{1992MNRAS.254..647I}, $26.2 \pm 5.4$ keV \citep{2009A&A...496..121B}, and 36.2 $_{-3.2}^{+3.5}$ keV \citep{2018MNRAS.474.1564W}. We estimated the largest value of the maximum plasma temperature at minimum, 53 $_{-7}^{+6}$ keV, possibly indicating that the plasma may undergo cooling only later, in the inter-outburst period. This may be assessed precisely only with grating observations with high resolution between the outbursts.
 
 Although the mass accretion rate increased towards maximum and decreased again during the decay, as expected, at minimum the value returned by our spectral fit is still about 6 times higher than the values obtained from fits to X-ray spectra taken later, in quiescence, by \citet{2018MNRAS.474.1564W} with a different model, indicating that GK Persei might be not have returned immediately to its pre-outburst state, although the unabsorbed flux, 
 $\sim 2.3 \times 10^{-11}$ erg cm$^{-2}$ s$^{-1}$, was already close to the quiescence values in the 0.3 -- 10 keV range of 2.7 $\times10^{-11}$ and 4.5 $\times10^{-11}$ erg cm$^{-2}$ s$^{-1}$ reported by \citet{1988MNRAS.231..783N} and of 2.3 $\times10^{-11}$ erg cm$^{-2}$ s$^{-1}$ reported by \citet{2009MNRAS.399.1167E}.


\begin{table*}
\centering
\begin{minipage}{158mm}
\caption[The best fitting model parameters of the {\sl Swift} XRT data]{The best fitting model parameters of the {\sl Swift} XRT data of GK Persei during the 2018 outburst. The model is \texttt{tbabs $\times$ (BB + pwab $\times$ (mkcflow + gaussian))}. 
The errors represent the 90\% confidence region for a single parameter.}
\scalebox{0.99}{
\begin{tabular}{llcccccc}
\hline
Component & Parameter						&  \multicolumn{6}{c}{Value} 					\\
		  & 								& epoch 1		  & epoch 2	                          & epoch 3                       & epoch 4                & epoch 5              & epoch 6             	\\
\hline
tbabs	  & nH$^{a,b}$ 						                 & 0.17			   & 0.17                                 & 0.17                          & 0.17                           & 0.17                             & 0.17             \\
\hline 
 		  & nH$_{\rm min}$$^c$  			          &76 $_{-9}^{+8}$               &2.2 $_{-0.40}^{+0.45}$               &46 $_{-10}^{+7}$                   &0.28 $_{-0.17}^{+0.18}$         &15 $_{-6.5}^{+6.9}$                &1.4 $_{-0.8}^{+0.8}$                      \\    
pwab	  & nH$_{\rm max}$$^b$				                  &23.9 $_{-0.3}^{+1.0}$ 	 &48.4 $_{-3}^{+4}$                    &55 $_{-5}^{+7}$                   &28 $_{-2}^{+2}$                 &18 $_{-2}^{+2}$                    &3.1 $_{-0.3}^{+0.4}$    	\\    
		  & $\beta$						  &0.56 $_{-0.08}^{+0.9}$	 &0.58 $_{-0.08}^{+0.10}$              &0.51 $_{-0.19}^{+0.20}$            &0.46 $_{-0.08}^{+0.9}$         &0.53 $_{-0.11}^{+0.13}$            &0.44 $_{-0.27}^{+0.32}$    	       \\
\hline   
mkcflow   & T$_{\rm high}$ (keV)			                  &25 $_{-4}^{+6}$	         &38  $_{-6}^{+6}$                     &41 $_{-10}^{+8}$                   &20 $_{-4}^{+5}$                 &39 $_{-8}^{+9}$                    &53 $_{-7}^{+6}$                 \\
		  & $\dot{m}$$^d$					  &12 $_{-2}^{+2}$               & 11 $_{-1}^{+2}$                     &15 $_{-4}^{+3}$                   &23 $_{-1}^{+1}$                 &7.2 $_{-0.9}^{+1.3}$                &1.0 $_{-0.2}^{+0.2}$                   \\ 
\hline
		  & E (keV)$^a$						  & 6.4     		         & 6.4                                 & 6.4                              & 6.4                            & 6.4                                & 6.4                    \\
Gaussian  & $\sigma$ (keV)$^a$ 				                  & 0.04		         & 0.04                                & 0.04                             & 0.04                           & 0.04                               & 0.04                       \\
\hline  
BB		  & T (eV) 						  &66 $_{-7}^{+8}$  	         & 75   $_{-7}^{+6}$                   &77 $_{-7}^{+6}$                   &83 $_{-10}^{+9}$                 &66 $_{-17}^{+26}$                   &89 $_{-42}^{+51}$                        \\
\hline    
Flux$_{\rm 0.3-2\ keV}$$^e$& abs.    		                          &1.9 $_{-0.2}^{+0.1}$	         &3.0  $_{-0.3}^{+0.3}$                &6.7 $_{-0.8}^{+0.9}$              &2.5 $_{-0.3}^{+0.4}$            &1.3 $_{-0.1}^{+0.2}$                &1.5 $_{-0.3}^{+0.2}$                  \\
						  & unabs. 		  &95  $_{-5}^{+7}$ 	         &141   $_{-10}^{+15}$                  &272 $_{-35}^{+30}$                &159 $_{-9}^{+10}$               &70 $_{-7}^{+8}$                     &11 $_{-1}^{+2}$                        \\
Flux$_{\rm 2-10\ keV}$$^e$ & abs.    		                          &94  $_{-5}^{+3}$	         &127  $_{-6}^{+7}$                    &125 $_{-12}^{+11}$                  &136 $_{-9}^{+6}$                &119 $_{-9}^{+6}$                    &21 $_{-3}^{+2}$                     \\
						  & unabs. 		  &142 $_{-6}^{+5}$	         & 294   $_{-20}^{+13}$                &283 $_{-21}^{+22}$                &223 $_{-10}^{+13}$              &153 $_{-14}^{+9}$                   &23 $_{-2}^{+2}$                    \\						  
Flux$_{\rm mkcflow}$$^e$& abs.    		                          &96 $_{-17}^{+19}$	         &121  $_{-13}^{+15}$                  &122 $_{-14}^{+14}$                &131 $_{-15}^{+14}$            &119 $_{-15}^{+17}$                    &22 $_{-6}^{+7}$                  \\
						  & unabs. 		  &219 $_{-33}^{+34}$	         & 407   $_{-31}^{+28}$                &416 $_{-40}^{+41}$                &367 $_{-22}^{+25}$              &210 $_{-19}^{+20}$                  &28 $_{-7}^{+8}$                       \\
Flux$_{\rm Gaussian}$$^f$& abs.    		                          &220 $_{-103}^{+98}$	         &140  $_{-85}^{+84}$                  &374 $_{-131}^{+126}$              &182 $_{-92}^{+85}$            &65 $_{-31}^{+36}$                     &24 $_{-12}^{+17}$                  \\
						  & unabs. 		  &300 $_{-126}^{+131}$	         & 323   $_{-165}^{+179}$              &835 $_{-304}^{+287}$              &289 $_{-138}^{+143}$              &78 $_{-36}^{+37}$                 &28 $_{-12}^{+17}$                       \\
Flux$_{\rm BB}$& abs.$^f$    		                                  &22 $_{-11}^{+12}$	         &24  $_{-12}^{+13}$                   &79 $_{-34}^{+32}$                 &13 $_{-6}^{+6}$               &5 $_{-3}^{+2}$                        &10 $_{-5}^{+5}$                  \\
						  & unabs.$^e$ 		  &12 $_{-2}^{+3}$	         & 13   $_{-2}^{+3}$                   &36 $_{-4}^{+5}$                   &10 $_{-2}^{+2}$               &3 $_{-1}^{+1}$                        &0.89 $_{-50}^{+52}$                        \\
\multicolumn{1}{l}{L$_{\rm BB}$ ($\times10^{33}$erg s$^{-1}$)}   &        &0.70 $_{-0.10}^{+0.10}$       &0.74  $_{-0.09}^{+0.10}$             &1.9 $_{-0.2}^{+0.3}$              &0.48 $_{-0.08}^{+0.10}$         &0.17 $_{-0.07}^{+0.07}$             &0.02 $_{-0.01}^{+0.01}$                     \\
\multicolumn{1}{l}{R$_{\rm BB}$$^g$ ($\times10^5$cm)}	    &             &11.5 $_{-1.2}^{+1.2}$         &11.9  $_{-0.7}^{+0.8}$               &19.8 $_{-1.1}^{+1.2}$             &9.1 $_{-0.8}^{+0.8}$            &4.8 $_{-0.8}^{+0.9}$                &1.9 $_{-0.8}^{+0.7}$                   \\			
\multicolumn{1}{l}{L$_{\rm 2-10\ keV}$ ($\times10^{33}$erg s$^{-1}$)} &   &3.3 $_{-0.2}^{+0.1}$	         &6.9 $_{-0.5}^{+0.3}$	               &6.6 $_{-0.5}^{+0.6}$              &5.2 $_{-0.2}^{+0.3}$            &3.5 $_{-0.2}^{+0.2}$                &0.52 $_{-0.05}^{+0.05}$ 		\\
\multicolumn{1}{l}{$\chi^2$}				   &              &0.92		                 & 0.93                                & 0.81	                          & 1.06                           & 0.88                               & 0.91                                           \\
\hline
\multicolumn{8}{p{.9\textwidth}}{{\bf Notes}: $^a$Frozen parameter. $^b$$\times$10$^{22}$ cm$^{-2}$. $^c$$\times$10$^{20}$ cm$^{-2}$.}\\
\multicolumn{8}{p{.9\textwidth}}{$^d$Mass accretion rate $\times$10$^{-10}$ M$_\odot$ yr$^{-1}$.
$^e$$\times10^{-12}$ erg cm$^{-2}$ s$^{-1}$. $^f$$\times10^{-14}$ erg cm$^{-2}$ s$^{-1}$.}\\
\multicolumn{8}{p{.9\textwidth}}{$^g$Radius of the emitting region. R$_{\rm BB}$ is a formal parameter, much smaller than the actual size of the optically thin emitting region. The distance is assumed to be 441 pc.}\\
\hline 
\end{tabular}
}
\label{tab:model parameters}
\end{minipage}
\end{table*}

\section{Conclusions}
GK Persei is the CV that ``has it all'', allowing us to study a vast range of phenomena in an object that is at relatively close distance. It is unusual, but it is also a ``Rosetta stone'' of WD binaries at the same time. By monitoring it, we acquire important information about fundamental physical parameters that allow us to construct physical models. With X-ray and UV observations of the 2018 eruption, we made some progress in understanding what parameters vary, and how, in the unusual, rather unique DN-like outburst of a magnetic CV that also undergoes classical nova outbursts. The following are the important conclusions we were able to draw from this recent eruption.

The WD spin period of $ 351.325 \pm 0.009$ s was 
still detected in the 2 -- 10 keV range
 in the 2018 outburst, but the modulation was less prominent than in the 2015 outburst. We found a direct correlation of the
 pulsed fraction with the mass accretion rate, at least as it can be estimated with spectral fits.

The spin modulation of the hard 
 X-ray flux (above 2 keV) indicates that the hard X-rays originate from the accretion columns that are funneled onto the poles, and that there is a larger shock height than in other IPs in which the hard X-ray flux is smaller and/or not much modulated with the WD spin. 
We confirmed that the spin modulation is not measurable 
 down to a level of about 9 \%, in the softest energy band, 0.3-2 keV. This is puzzling, because in many IPs the modulation seems to be caused by absorption, probably due to the ``accretion curtain'', and the effect of the column density is of course larger at lower energy \citep[e.g.][]{Rawat2022}. The absence of pulsations in soft X-rays shows that the soft, blackbody-like component does not originate from a small hot spot close to the poles, like in other accreting magnetic WDs. Yet, the low luminosity of this component implies a very small size of the emitting region. We suggested that we are observing diffuse thermal plasma that is not on the WD surface, possibly originating in a wind from the disk. If there is a second component at plasma temperature around 100 eV or less, in a CCD-like detector it cannot be differentiated from a blackbody. Since we also know from the high resolution spectra that there is very little or no continuum flux in the 0.4 -- 2 keV range (6 -- 30 \AA), and only emission lines are detected, which do {\it not} show a pulsed fraction \citep{2017MNRAS.469..476Z}, it is likely that most of the soft X-ray flux does not originate in the accreted material, but rather in diffuse plasma: we suggest a disk wind, or similar mass loss phenomenon.

The spin modulation was undetectable in the UV in the
2018 outburst, like in the previous, more common outbursts of about the same
 amplitude in optical and X-rays, while the modulation was present in the UV in 
a rarer, lower amplitude outburst that also had, unusually, several optical peaks. The physical meaning of this difference may only be explored in a new outburst with
 low amplitude. 
 
For the first time, we had the possibility to follow
 the X-ray and UV light curves also during the decay of luminosity, and not only the rise to maximum. With our spectral fits, we were able to confirm that the mass accretion rate increased during the rise to maximum and decreased in the decay towards minimum, although we found evidence that it was still enhanced at minimum with respect to the inter-outburst value derived many months after the DN eruption. We could not estimate the maximum temperature with precision, but our fits indicate that it may have varied irregularly during the outburst, rather than being closely correlated with the mass accretion rate onto the WD.
 There may also be a continuous cooling over the first inter-outburst months.

Finally, we also found evidence that the mass accretion rate varies significantly in the different DN-like outbursts of GK Persei. The values we obtained in the spectral fits of the 2018 outburst are about an order of
 magnitude lower than those derived for 2015 outburst by \citet{2017MNRAS.469..476Z}. 

\section*{Data Availability}
The data analyzed in this article are all available in the HEASARC archive of NASA at the following URL: \url{https://heasarc.gsfc.nasa.gov/db-perl/W3Browse/w3browse.pl}

\section*{Acknowledgements}
We thank the anonymous referees for their constructive comments and suggestions, which helped us to improve the scientific content of this paper. We acknowledge with thanks the observers worldwide for contributing their data to the AAVSO. This work made use of data supplied by the UK Swift Science Data Centre at the University of Leicester. Song-Peng Pei is supported by the High-level Talents Research Start-up Fund Project of Liupanshui Normal University (LPSSYKYJJ202208), Science Research Project of University (Youth Project) of the Department of Education of Guizhou Province (QJJ[2022]348), the Science and Technology Foundation of Guizhou Province (QKHJC-ZK[2023]442) and the Discipline-Team of Liupanshui Normal University (LPSSY2023XKTD11).




\bibliographystyle{mnras}
\bibliography{GKPer} 




\bsp	
\label{lastpage}
\end{document}